\documentclass[
  journal=largetwo,
  manuscript=article-type,
  year=2025
]{cup-journal}

\usepackage{amsmath}
\usepackage[nopatch]{microtype}
\usepackage{booktabs}
\usepackage{url}
\usepackage[normalem]{ulem}
\usepackage{xcolor}
\usepackage{pdflscape}
\usepackage[authordate-trad,noibid,backend=biber,natbib]{biblatex-chicago}

\title{Breakthrough Listen: A Technosignature Search Around 27 Eclipsing Exoplanets Selected from the \textit{Transiting Exoplanet Survey Satellite} Catalogue}

\author{R. Barrett}
\affiliation{University of Southern Queensland, Brisbane, 4300, QLD, Australia}
\email[R. Barrett]{rebeccabarrett95@hotmail.com}

\author{C.D. Tremblay}
\affiliation{SETI Institute, 339 Bernardo Ave, Suite 200, Mountain View, CA 94043, USA}
\alsoaffiliation{Berkeley SETI Research Center, University of California, Berkeley, CA 94720, USA}
\alsoaffiliation{CSIRO Astronomy and Space Science, PO Box 1130, Bentley WA 6102, Australia}

\author{B. Addison}
\affiliation{University of Southern Queensland, Centre for Astrophysics, West Street, Toowoomba, QLD 4350 Australia}
\alsoaffiliation{Swinburne University of Technology, Centre for Astrophysics and Supercomputing, John Street, Hawthorn, VIC 3122, Australia}

\author{D.C. Price}
\affiliation{International Centre for Radio Astronomy Research, Curtin University, Bentley, WA 6102, Australia}
\alsoaffiliation{SKA Observatory, Science Operations Centre, Kensington,
WA 6151, Australia}

\author{J. A. Green}
\affiliation{SKA Observatory, Science Operations Centre, Kensington,
WA 6151, Australia}


\begin{document}

\begin{abstract}
Here we analyse the archival data for a set of 27 \textit{Transiting Exoplanet Survey Satellite} (\textit{TESS}) Targets of Interest (TOIs) in search for artificially generated radio signals, or `technosignatures', interrupted by occultation. Exoplanetary eclipses are notable events to observe in the search for technosignatures, as they mark the geometrical alignment of the target, its host star, and Earth. During an eclipse event, any signal emanating from the target of interest should cease for the duration of the eclipse, and resume after the line-of-sight has been restored. Target observations were made by Breakthrough Listen (BL) using Murriyang, the CSIRO Parkes 64-m radio telescope, coupled with the Ultra-wide Low frequency (UWL) receiver covering a continuous range of frequencies spanning 704--4032 MHz inclusive. Each target was observed in a pattern consisting of six back-to-back 5-minute source and reference sky positions for comparison during data analysis. We performed a Doppler search for narrowband signals with a minimum signal-to-noise (S/N) ratio of 10, a minimum drift rate of $\pm0.1$\,Hz/s, and a maximum drift rate of $\pm4.0$\,Hz/s using the \textsc{turboseti} pipeline. In the analysis of 1,954,880 signals, 14,639 passed automated radio interference filters where each event was presented as a set of stacked dynamic spectra. Despite manually inspecting each diagram for a signal of interest, all events were attributed to terrestrial radio frequency interference (RFI).
\end{abstract}

\section{Introduction} \label{sec:intro}

Rapid cultural, scientific, and technological advancements over the last millennium have expanded our collective anthropocentric perspective to ponder our place among the stars, and the potential reality of a Universe teeming with intelligent life.

The Search for Extraterrestrial Intelligence (SETI), is an investigation into the presence of intelligent life beyond Earth via the search for signals from advanced technology (technosignatures). Experimental SETI began in 1959 with a proposed search for interstellar communications in the form of electromagnetic radiation \autocite{Cocconi}. In the decades that have followed, SETI science has secured a place in the scientific research community; and while potential means of viable interstellar communication have expanded in recent years to include methods such as gravitational wave modulation \citep{Sellers_2022}, electromagnetic radiation --- particularly in the radio portion of the electromagnetic spectrum --- remains at the forefront of SETI investigations (i.e. \citealt{Wright_2022,Haqq-Misra_2022,Sheikh2025}). Radio signals are relatively low in energy and cost-effective to produce and experience minimal scattering as they travel through planetary atmospheres and the interstellar medium \citep{Cordes_1991,Tremblay_2022}. The frequency band of radio data used in this investigation was carefully selected from a large set of radio observations made by $Breakthrough$ $Listen$\footnote{\url{https://www.physics.ox.ac.uk/research/group/breakthrough-listen}} and includes a selection of known and candidate exoplanets.

The ``Breakthrough Listen (BL) Initiative" \citep{Worden_2017} is an undertaking to rigorously search for technologically-capable life beyond Earth via technosignature detection. In its initial years, BL used Murriyang, the Parkes \citep{Price_2020} 64 m telescope, the Robert C. Byrd Green Bank telescope (GBT; \citealt{enriquez17}), and the Automated Planet Finder (APF; \citealt{Zuckerman_2023}), to survey nearby stars \citep{Isaacson_2017} for radio emission and for narrowband laser lines. These surveys have been augmented to search for other objects, such as exoplanets (i.e. \cite{BLSIL2022, BLSIL2022a, Barrett_2023}), a broad sampling of astrophysical objects \citep{Lacki_2021}, stars within the FoV (i.e. \cite{BLSIL2021d, Perez2022}), and galaxies \citep{Choza}.

Our presence on Earth is the ultimate piece of evidence that life can arise, evolve, and flourish in the environmental conditions that such a terrestrial planet provides. While it appears that terrestrial exoplanets are justifiably the target of choice in our search for life beyond Earth \citep{Kepler-442b}, the search for technosignatures as a sign of intelligent life is, in effect, a search for transmitter technology: technology that need not be located on or around traditionally habitable, nor necessarily inhabited, astronomical bodies.

While this study does focus particularly on exoplanets (both candidate and confirmed), the physical characteristics of each target was not considered in the target selection process, and thus includes a range of non-terrestrial bodies. Each target was selected on the basis of the predicted instance of secondary eclipse events during the observations. Information on the physical characteristics and epochs of observational events for each target is provided in Table \ref{tab:TSI}\footnote{The physical characteristics for each target system were obtained as of 2025 January 21 through the Exoplanet Follow-up Observing Program (ExoFOP), accessible through: \url{https://exofop.ipac.caltech.edu/tess/}.}

The idea of observing occultations to discover and confirm targets for SETI technosignature searches has gained in popularity over the last decade, with a particular focus on planet-planet occultation and signal spillover \citep{Tusay_2024, Sneed2023}. Here, however, we explore planet-star occultation in a rare study to observe the potential signal drop-off at the predicted time of eclipse. This work is based on the Master's thesis by \cite{Barrett_2023} in which the first limits using TESS targets were achieved. In \cite{Sheikh_2023_Kepler}, they searched for technosignatures toward 12 exoplanets discovered by the $Kepler$ space mission with the GBT. As a variation of this method, \cite{Tusay_2024} demonstrated a method to find signals during planetary-planetary occultation. However, this is the first work on to provide limits towards TESS Targets of Interest (TOIs) during planetary to star occultation.

Observational data is publicly available in the TESS Input Catalogue (\textit{TIC})\footnote{Accessible through MAST: \url{https://mast.stsci.edu/portal/Mashup/Clients/Mast/Portal.html}.}, with an ever-expanding list of confirmed and candidate targets available through the NASA Exoplanet Archive\footnote{The full list of \textit{TESS} Project Candidates is accessible through: \url{https://exoplanetarchive.ipac.caltech.edu/cgi-bin/TblView/nph-tblView?app=ExoTbls&config=TOI}.}. Data from both sources were used in the cross-referencing and determination of the final list of 27 transiting exoplanets analysed in this study.

Here, we report on a search through archival BL data from Murriyang, the Parkes 64-m radio telescope, for narrowband drifting technosignatures interrupted by occultation around 27 TESS TOIs selected from the TIC.

\section{Target Selection}
\label{sec:targetselection}
The target selection process that resulted in the final list of 27 eclipsing exoplanets involved filtering through all Breakthrough Listen Parkes observations between 2018 January 24 and 2022 June 11 inclusive for those stellar targets with an exoplanet that passed either the calculated point of secondary ingress or egress during the 30 minute observing window. This allowed for the exploration of the idea that if a technological signal were detected, it could potentially be localised to a particular exoplanet due to the cessation of the signal that would occur as the exoplanet passed behind the host star in eclipse or secondary transit.

To begin this reduction process, the full list of targets associated with each observing session was converted to a common catalogue identifier and cross-referenced with the host stars of confirmed transiting exoplanets in the \textit{TIC}. We then used the NASA Exoplanet Archive Transit Ephemeris Service\footnote{Ingress and egress predictions were calculated using the NASA Exoplanet Archive Transit and Ephemeris Service available at: \url{https://exoplanetarchive.ipac.caltech.edu/cgi-bin/TransitView/nph-visibletbls?dataset=transits}} to calculate the predicted eclipse ingress and egress times for each target within $\pm30$ minutes of each observation. This resulted in our list of 27 exoplanets that underwent secondary ingress or egress in a total of 25 observing sessions\footnote{Two stellar targets were observed twice, in different epochs.} or 150 $\times$ 5-minute cadence target\_S and target\_R observations.

\section{Observations}
\label{sec:obs}
This study uses a subset of archival data acquired by BL using Murriyang, the CSIRO Parkes 64-m radio telescope, located in New South Wales, Australia (32.9986 $\deg$ S, 148.2621 $\deg$ E). Each target was allocated 30 minutes of observation time which was divided into six consecutive five-minute observations nodding back and forth between the target source (target\_S) and a reference target (target\_R) offset by 0.5 degrees. This observation strategy is an effective method to identify local sources of radio frequency interference (RFI), since local RFI should persist in both target\_S and target\_R frames \citep{BLSIL2019f}. For more details on the observation setup and configuration, see \cite{Price_2020}.

Between the dates 2018 January 24 and 2022 June 11 inclusive, BL used the the Ultra-Wideband Low frequency (UWL;\citealt{UWL}) receiver to capture raw voltage data over a continuous range of frequencies spanning 704 -- 4032\,MHz. The full bandwidth (3200\,MHz) of observational data from each source and reference observation was evenly split into 25 $\times$ 128\,MHz segments and shared among the 27 compute nodes (26 and 1 spare) that make up the BL Parkes back-end \citep{BLSIL2018d, BLSIL2021c}. Here, the raw voltage data from the UWL receiver was processed into high-frequency resolution time-series data files using the BL-Parkes Data Recorder signal processing system.

\section{Data Analysis}
\label{sec:DA}
To complete the data analysis, we used three high-power compute nodes at the Breakthrough Listen Data Centre located at the University of California, Berkeley.
As previously mentioned, the BL Parkes back-end recorded the raw voltage data for each cadence in sub-bands of 128\,MHz spanning the range of frequencies associated with the UWL receiver (704 -- 4032 MHz). Unfortunately, the data associated with sub-band 3648 -- 3776 MHz failed to record during observations, leading to a gap of 128 MHz in the analysed spectrum, impacting only 4\% of the data.

\begin{figure*}
\includegraphics[width=0.8\textwidth]{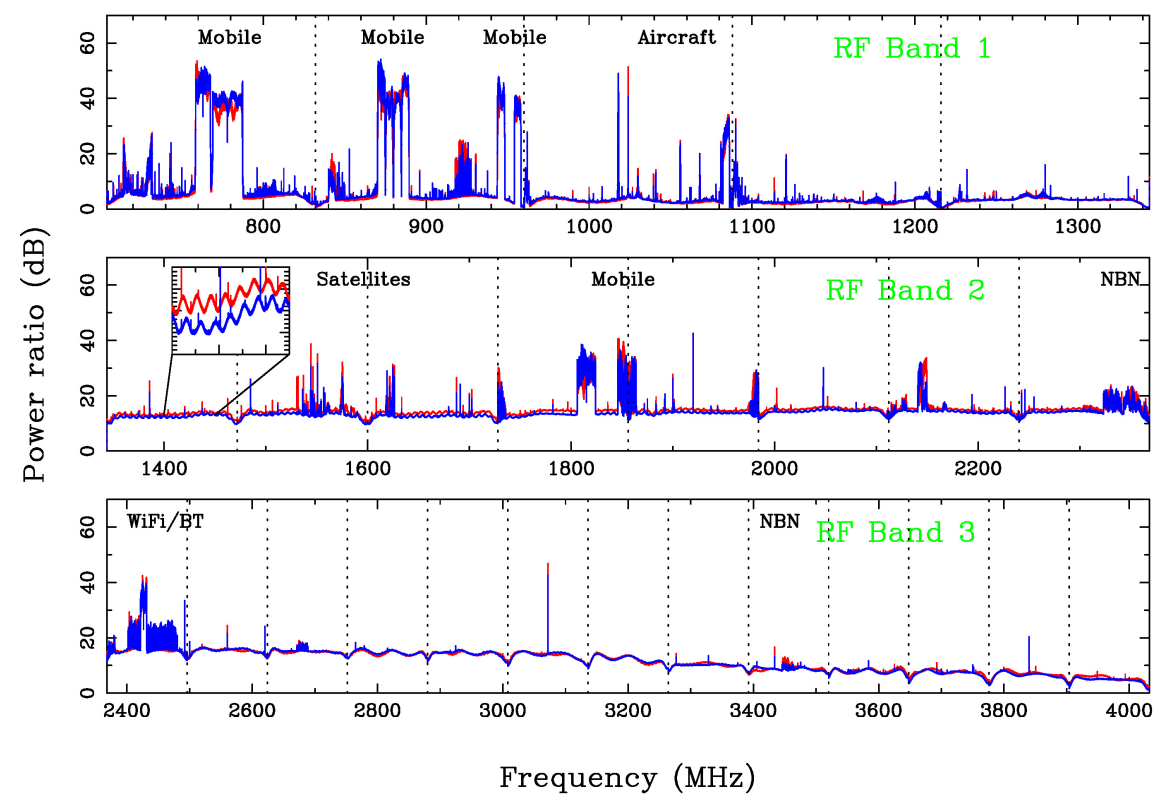}
\caption{Spectra showing local sources of RFI that consistently affect the Parkes UWL bandpass. The UWL receiver records two polarisations
which are represented by the colours red and blue, respectively, in this diagram. Reproduced with permission from \citep{Hobbs}.}
\label{fig:UWLRFI}
\end{figure*}

We did not filter any regions of radio frequency interference (RFI, Figure \ref{fig:UWLRFI}) out before the analysis\footnote{We anticipate that RFI impacts approximately 25\% of the total UWL receiver band (calculated by averaging the remaining 'band use' data published in Table 1 of \citet{Hobbs}), resulting in a large fraction of false positives described in this region. We are working on improving techniques to eliminate known persistent RFI prior to analysis that will be used in future studies. The BL team, in general, is also designing improved algorithms that remove the RFI and iterates the search to find weaker signals. However, these tools are still in active development, and were not suitable for use when this manuscript was written.}. Therefore, a total of 3,650 high frequency resolution ($\sim$ 2 Hz) \textsc{HDF5} data files amounting 20\,TB were accessed and analysed using BL's \textsc{turboseti}\footnote{\url{https://github.com/UCBerkeleySETI/turbo_seti}} pipeline \citep{turboSETIPaper}. The \textsc{turboseti} pipeline is a package of \textsc{python} scripts designed specifically to search high-resolution data files for continuous, drifting, narrowband signals using the Taylor-tree de-doppler algorithm \citep{Taylor1974}.

First, the \textsc{FindDoppler} class was invoked, allowing the specification of search parameters relating to the: minimum and maximum signal doppler drift rate (change in signal frequency due to relative radial motion; \citealt{Li_2022}), the minimum signal to background noise ratio (S/N), and the number of coarse channels (frequency sub-bands) in which to divide and search each \textsc{HDF5} data file. We adopt a minimum signal doppler drift rate of $\pm$ 0.1 Hz/s, accounting for the fractional drift rate of Earth \citep{DriftRate}, and a maximum drift rate of $\pm$ 4 Hz/s, in line with \citet{sheikh2019}. A minimum S/N value of 10 was selected following \citealt{BLSIL2020a}. Recent work evaluating the performance of \textsc{turboseti} supports the continued use of this value in future studies, indicating that the background noise is non-Gaussian and thus exhibits a higher false positive rate than anticipated (\citealt{Choza}, Tremblay et al. in prep). This value represents a trade-off between sensitivity and the number of false positives due to RFI combined with changes in sky temperature and side-lobe behaviour between target source and reference pointing's. Finally, the number of coarse channels was set to a value of 32, allowing the script to search for hits in further subdivisions of 6\,MHz.

For each file that returned a successful \textsc{FindDoppler} check, two new data files were created containing detailed information about each flagged signal. The `\textsc{find\_event\_pipeline}' class was then invoked to filter for any signals that appeared in all target\_S and no target\_R observations for each set of associated cadences. For each of these flagged events, the `\textsc{plot\_event}' pipeline then created individual dynamic spectra for final visual inspection.

\section{Results}
\label{sec:res}

\subsection{Doppler Search}
\label{subsec:DS}
Data analysis using the \textsc{turboseti} `\textsc{FindDoppler}' class returned a total of 1,954,880 hits, of which 14,639 passed the `\textsc{find\_event\_pipeline}' parameters and were flagged as possible events. This resulted in the `\textsc{plot\_event}' creation and subsequent visual inspection of 14,639 waterfall diagrams.

\begin{figure}
\includegraphics[width=1.0\textwidth]{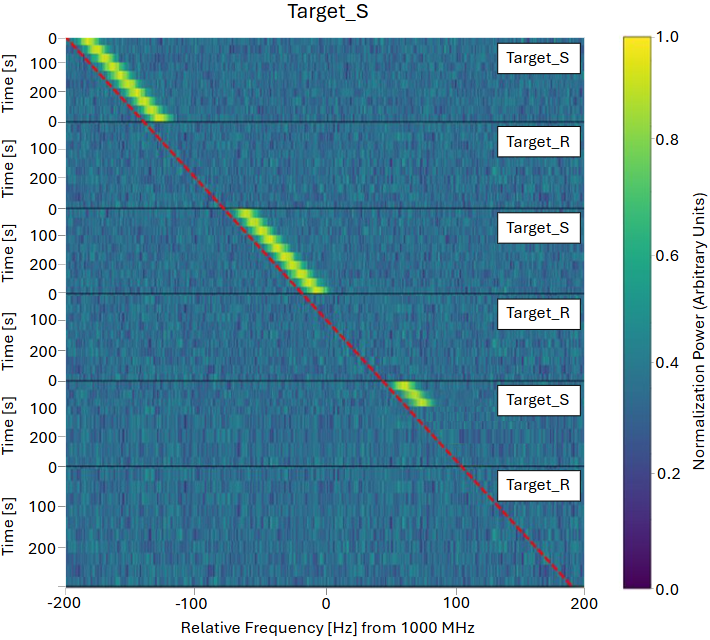}
\caption{An idealised example of a continuous narrowband drifting technosignature interrupted by occultation. In this scenario, a transmitter in the vicinity of the target exoplanet emits a continuous narrowband signal until the fifth frame (second from the bottom), where the signal drops off at the predicted time of eclipse. The signal is strong, matches the predicted drift rate for the target (red dashed line), is present in all target\_S frames, and absent in all target\_R frames, near eliminating the possibility of an RFI source. This illustration was created in partial using \textsc{setigen} (\citealt{brzycki2022setigen}).}
\label{fig:SETI_Gen}
\end{figure}

\begin{figure}
\includegraphics[width=0.9\textwidth]{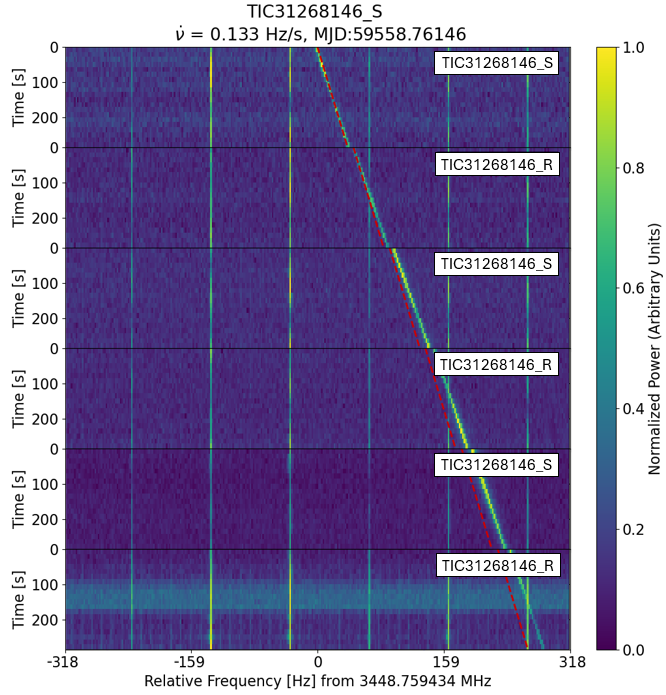}
\caption{This diagram features one drifting narrowband event signal centred around 3448.76 MHz accompanied by a set of six evenly-spaced fixed-frequency background signals. The event signal aligns well with the predicted drift rate (red dashed line), yet all signals are present in both target source (TIC31268146\_S) and reference target (TIC31268146\_R) frames, indicating RFI origin. Whilst a drifting signal would suggest emission from a moving target, in Figure \ref{fig:UWLRFI} would suggest it likely that all signals in this figure relate to the frequencies emitted by Australia's National Broadband Network (NBN), or other networks operating at similar frequencies.}
\label{fig:WF1}
\end{figure}

\begin{figure}[h]
\includegraphics[width=1.0\linewidth]{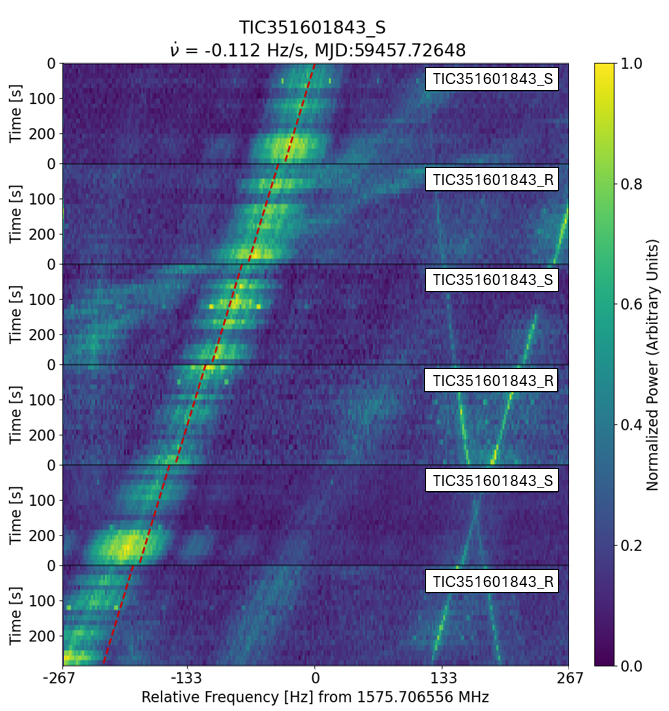}
\caption{This diagram features an assortment of signals with varying strengths, drift rates, and drift rate evolutions centered around 1575.71 MHz. The event signal in this figure pairs well with its predicted drift rate, yet it is once again present in both target\_S and target\_R frames, indicating RFI. It is worth noting that while each diagram has a focus on one particular event signal, prominent drifting background signals that pass \textsc{turboseti} search parameters will also have been flagged as separate events and are analysed separately. According to Figure \ref{fig:UWLRFI} and the ATNF RFI Frequency List, this central frequency lies within the realm of satellite communications and signals sent to and from mobile devices.}
\label{fig:WF2}
\end{figure}

\begin{figure}[h]
\includegraphics[width=1.0\linewidth]{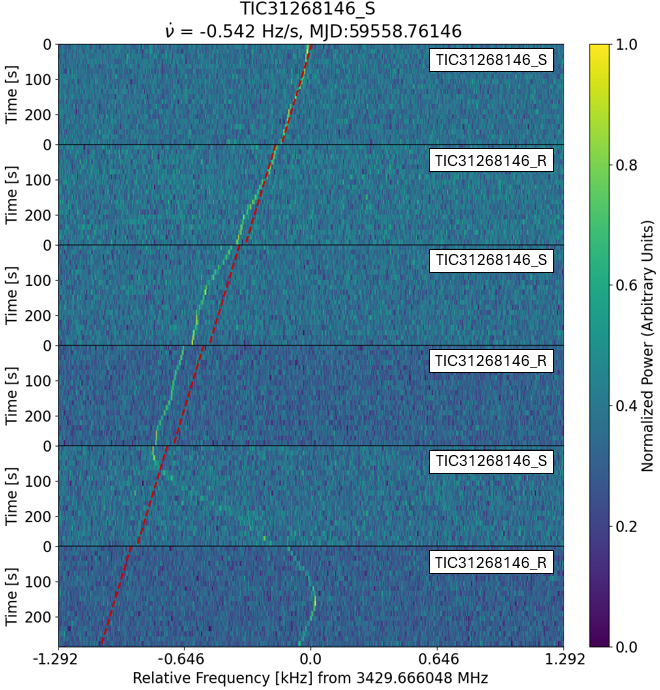}
\caption{This final diagram features a peculiar signal starting at a central frequency of 3429.67 MHz. \textsc{turboseti} is limited to linear drift rate predictions, and this signal quite clearly deviates from \textsc{turboseti} estimates. Figure \ref{fig:UWLRFI} would suggest that this signal lies well within the region of the NBN, yet the erratic drift rate evolution and singular nature of this signal is likely due to emission from a local moving source, such as a small plane flying over the telescope's field of view during observation.}
\label{fig:WF3}
\end{figure}

Figure \ref{fig:SETI_Gen} provides an idealised example of a continuous narrowband drifting signal interrupted by occultation\footnote{Created in partial using the \textsc{setigen} python package available at \url{https://github.com/bbrzycki/setigen}.}. Figures \ref{fig:WF1}, \ref{fig:WF2}, and \ref{fig:WF3} are direct representations of some of the types of signals and signal groupings that were present throughout inspection. While the event signal in each of these figures aligns relatively well with the predicted signal drift rate (red dashed line), each was easily identified as RFI due to being present in both target\_S (on source) and target\_R (off source) observations. Further investigation suggests emission from known sources of local RFI operating at similar frequencies, such as those listed in CSIRO's Australian Telescope National Facility RFI Frequency List\footnote{Publicly available at: \url{https://www.parkes.atnf.csiro.au/observing/rfi/parkes_rfi_survey/frequency_list.html}} and Figure \ref{fig:UWLRFI}. For each event signal that passed \textsc{turboseti} search parameters, visual inspection revealed that all were attributable to RFI.

\subsection{Figures of Merit}
\label{subsec:FoM}
In line with past SETI technosignature publications \citep{BLSIL2017b, Margot}, one metric along with one Figure of Merit (FM) has been calculated to compare the efficiencies and effectiveness of this study to those of the past and future.

\subsubsection{Equivalent Isotropic Radiated Power (EIRP) Metric}
\label{subsubsec:EIRP}
The first of these values is the minimum Equivalent Isotropic Radiated Power ($EIRP_{min}$), which describes the minimum amount of power ($W$) required to transmit a detectable omni-directional signal of a particular central frequency from a transmitter with the equivalent size and capabilities of the receiving telescope \citep{BLSIL2017b}. A minimum EIRP was calculated for each observed exoplanet using the following equation \citep{BLSIL2017b}:

\begin{equation}
    EIRP_{min} = 4\pi{}d^{2}F_{min} ,
\end{equation}

\noindent where $d$ is the distance to the target ($m$), and $F_{min}$ is the minimum detectable flux ($W/m^{2}$). Following \citet{BLSIL2020a}, the minimum detectable flux for a narrowband signal was calculated using:

\begin{equation}
    F_{min} = {S_{min}}{\delta\nu} ,
\end{equation}

\noindent where $S_{min}$ is the minimum flux density ($W/m^{2}/Hz$), and $\delta\nu$ is the bandwidth of the transmitting signal ($Hz$). Following \citealt{enriquez17}, we set $\delta\nu$ to unity. 

Finally, the minimum flux density for a narrowband signal is given by \citep{BLSIL2017b}:

\begin{equation}
    S_{min} = {S/N_{min}}SEFD\sqrt{\frac{B}{n_{pol}\tau_{obs}}} ,
\end{equation}

where $B$ is the receiving channel bandwidth ($Hz$), $S/N_{min}$ is the minimum signal to noise ratio,  $n_{pol}$ is the number of polarisations, $\tau_{obs}$ is the observing time (s), and $SEFD$ is the System Equivalent Flux Density (SEFD). The SEFD describes the sensitivity of the radio observation as a function of system noise and the effective collecting area of the receiving telescope. \citet{UWL} list a set of median SEFD values for each sub-band of the Parkes UWL receiver, with a minimum value of 33\,Jy at a central frequency of 1664\,MHz --- this is the value adopted for the SEFD in this study. As for the other parameters, $S/N_{min}$ was set to a value of 10 (the S/N ratio used during \textsc{turboseti} analysis), $n_{pol}$ to 2 (as the UWL receiver records over two polarisation channels)\footnote{Due to an equipment failure spanning dates 2019 January 05 (58604 MJD) and 2021 June 06 (59386 MJD) inclusive, the UWL receiver was only able to record over one polarisation. This is accurately reflected in the calculated $EIRP_{min}$ values in Table \ref{tab:TSI}, where $n_{pol}$ was set to 1.}, $\tau_{obs}$ to 300 seconds (equivalent to an observing cadence of 5-minutes), and $B$ to 2 Hz.

The minimum EIRP value calculated for each target system is listed in Table \ref{tab:TSI}, encompassing values between $9.46\times10^{11}$ $W$ and $1.27\times10^{15}$ $W$, inclusive. While recording over either one or two polarisations has not dramatically affected EIRP calculations, it should be noted that in the case that the transmitted signal were circularly or linearly-polarised, recording just one polarisation could indeed capture either half, all, or none of the transmitted signal depending on its match to the recorded polarisation.

\subsubsection{Continuous Waveform Transmitter Figure of Merit (CWTFM)}
\label{subsubsec:CWTFM}
The CWTFM is a metric designed by BL to standardise and compare the efficacy of past and future BL SETI surveys with respect to the volume of space searched in each project \citep{Margot}. This study uses archival data from a targeted survey, with a focus on a single target star per pointing. Having been selected by an observed event, rather than by parameters such as proximity, the targets analysed in this study are spread over a wide range of coordinates in the galactic southern hemisphere. With that noted, it is worth calculating such a value to compare to relevant future studies with a focus on analyzing specific targets. The CWTFM in this study was calculated as follows \citep{BLSIL2017b}:

\begin{equation}
    CWTFM = \zeta_{AO}\frac{EIRP_{min}}{{N_{stars}\nu_{rel}}},
\end{equation}

\noindent where $N_{stars}$ is the total number of stars --- equal to the number of pointings ($N_{pointings}$) multiplied by the number of stars ($n_{stars}$) per pointing; $\nu_{rel}$ is the fractional bandwidth --- equal to the total bandwidth ($\Delta\nu_{tot}$) divided by the central frequency ($\nu_{c}$); and $\zeta_{AO}$ is a normalization factor based on the now retired Arecibo Observatory, providing a standard by which each study can be compared. 

A value of $5\times10^{-11} W^{-1}$ was calculated for $\zeta_{AO}$ using the EIRP, $\nu_{rel}$, $N_{stars}$, and CWTFM values given in \citet{BLSIL2017b}. $EIRP_{min}$ was set to the largest value calculated in the results of this study, $1.27\times10^{15}$; $\nu_{rel}$ was set to a value of 1.9231 using a $\Delta\nu_{tot}$ value of 3200 MHz and a $\nu_{mid}$ value of 1664 MHz. Finally, $N_{stars}$ was set to 22 with the exclusion of TIC251852984, as the data for each of the 27 target exoplanets was collected from 23 telescope pointings, each with a focus on 1 host star. This set of values give a CWTFM result of $1496.48$.

\section{Conclusions}
\label{sec:conc}
Here we have performed a technosignature search towards 27 exoplanets that underwent occultation during observations made by BL with Murriyang, the CSIRO Parkes Radio telescope. A search for technosignatures could be enhanced by observing the ingress and/or egress points of an eclipsing exoplanet, as any continuous or rhythmic signal emission would be interrupted in the process, and the target could be easily located. The archival data for each target was searched for continuous or interrupted drifting narrowband signals spanning frequencies 704 -- 4032 MHz inclusive using BL's \textsc{turboseti} pipeline, followed by visual inspection. While no technosignatures were detected, our $EIRP_min$ calculations suggest that if our targets had the 20\,TW transmission capabilities of the late Arecibo radio telescope, 59.3\% would be capable of sending a signal powerful enough to be detectable from Earth. Further, a 20\,TW signal sent from an exoplanet orbiting the closest star to Earth in this study --- TIC 370133522 at 20.37\,pc --- would be detectable just above the 3$\sigma$ level. These are the first statistical limits on hypothesised technosignatures during exoplanetary eclipses for TESS targets observable in the southern hemisphere.

\subsection{Future Work}
\label{subsec:fw}
Additional observations of these sources at a broader range of frequencies, lower S/N ratios, and broader drift rate parameters in the future should be considered. The full list of targets assessed in this study could be re-observed and re-analysed to account for the data that failed to record over two polarisations, and the incomplete recording of frequencies in the 3648 -- 3776 MHz sub-band. The idea of signal interruption due to exoplanet occultation should be further explored with regard to signal of interest verification and perhaps as a means for detecting weak signals via the stacking of observational data at the point of secondary ingress and egress. Such image stacking should, in theory, amplify any consistent signals of interest and reduce the appearance of inconsistent RFI. Future studies should also consider that some of the more stringent filtering options available when using \textsc{turboseti} to automatically filter out potential candidate signals that do not appear in all target\_S cadences due to signal occultation. Next generation software is under development that will aid in setting better limits in future technosignature searches \citep{Ben2025,Ken2023}.

\begin{acknowledgement}
We thank M. Lebofsky at Breakthrough Listen for assisting us in obtaining the archival data associated with this project. Murriyang, the Parkes radio telescope, is part of the Australia Telescope National Facility (https://ror.org/05qajvd42) which is funded by the Australian Government for operation as a National Facility managed by CSIRO. We acknowledge the Wiradjuri people as the Traditional Owners of the Observatory site. We acknowledge additional support from other donors including the Breakthrough Prize Foundation under the auspices of Breakthrough Listen. This research has made use of the NASA Exoplanet Archive, which is operated by the California Institute of Technology, under contract with the National Aeronautics and Space Administration under the Exoplanet Exploration Program.
\end{acknowledgement}

\paragraph{Data Availability Statement}
All software used in this publication are available on GitHub and are referenced throughout the paper. The observations from the Parkes Telescope were collected during time allocated by Breakthrough Listen and can be accessed through \url{http://seti.berkeley.edu/opendata}. If particular data or computing resources are needed, they can be provided by direct enquiry to the Breakthrough Listen team. 

All other data that are publicly available through individual archives as stated within the relevant sections, including the NASA Exoplanet Archive (\url{https://exoplanetarchive.ipac.caltech.edu/}), the ExoFOP (\url{https://exofop.ipac.caltech.edu/tess/}), and the TESS target catalogue
(\url{https://tess.mit.edu/science/tess-input-catalogue/}). 

\printbibliography

\begin{landscape}
\begin{table}[hbt]
    \begin{threeparttable}
    \caption{Target System Information, Observation \& Event Epochs} 
    \label{tab:TSI}
        \begin{tabular}{lllllllllllll}
        \toprule
        \headrow Stellar Target ID & Class & Dist. (pc) & RA (h/m/s) & DEC (d/m/s) & Exoplanet ID & M ($M_{J}$) & R ($R_{J}$) & P (d) & a (AU) & Obs. Epoch (JD)\tnote{a} & Event Epoch (JD) & EIRP ($W$)\\
        \midrule
        TIC 118327533 & F & 745.22 & 00h41m22.68s & -37d20m34.80s & TOI 346.01 &  & 2.68 & 16.56 &  & 2459675.646 & 2459675.650 & $1.27\times10^{15}$ \\
        
        TIC 130924120 & G & 60.34 & 12h31m58.76s & -35d33m16.03s & TOI-757 b & 0.03 & 0.22 & 17.47 & 0.12 & 2459244.299 & 2459244.314 & $1.17\times10^{13}$* \\
        
        TIC 134200185 & K & 47.39 & 07h06m14.18s & -47d35m16.14s & TOI-500 b & 0.01 & 0.10 & 0.55 & 0.01 & 2459079.278 & 2459079.291 & $7.24\times10^{12}$* \\
        
        TIC 183985250 & G & 80.44 & 23h54m40.53s & -37d37m41.61s & LTT 9779 b & 0.09 & 0.42 & 0.79 & 0.02 & 2458894.800 & 2458894.818 & $2.09\times10^{13}$*\\
        
        TIC 220016044 & F & 480.32 & 02h09m17.25s & -48d29m05.06s & TOI 357.01 &  & 0.59 & 0.94 &  & 2459047.463 & 2459047.479 & $7.44\times10^{14}$* \\
        
        TIC 230982885 & G & 151.07 & 01h38m25.23s & -55d46m19.20s & WASP-97 b & 1.36 & 1.14 & 2.07 & 0.03 & 2458903.774 & 2458903.789 & $7.36\times10^{13}$* \\
        
        TIC 251852984\tnote{b} & K & 315.14 & 00h32m48.66s & -32d03m33.90s & TIC 323.01 &  & 0.57 & 2.77 &  & 2459047.230 & 2459047.248 & \\
        
        TIC 260304296 & F & 184.53 & 06h18m27.73s & -57d00m48.60s & TOI 187.01 &  & 1.72 & 0.51 &  & 2459310.689 & 2459310.696 & $1.10\times10^{14}$* \\
        
        TIC 260647166\tnote{c} & G & 64.60 & 12h26m17.78s & -51d21m46.99s & HD 108236 b & 0.01 & 0.14 & 3.80 & 0.05 & 2459090.461 & 2459090.443 & $1.35\times10^{13}$* \\
        
        &  &  &  &  & HD 108236 c & 0.02 & 0.19 & 6.20 & 0.06 &  &  & \\
        
        &  &  &  &  & HD 108236 d & 0.02 & 0.23 & 14.18 & 0.11 &  &  & \\
        
        &  &  &  &  & HD 108236 e & 0.03 & 0.28 & 19.59 & 0.14 &  &  & \\ 
        
        TIC 280206394 & F & 141.81 & 09h36m28.65s & -50d27m47.13s & TOI-677 b & 1.24 & 1.17 & 11.24 & 0.10 & 2459564.435 & 2459564.440 & $4.59\times10^{13}$ \\
        
        TIC 299799658 & G-K & 82.17 & 02h32m29.03s & -78d01m25.26s & TOI 1062.01 &  & 0.21 & 4.12 &  & 2459236.870 & 2459236.877 & $2.18\times10^{13}$* \\
        
        TIC 31268146 & G-K & 219.48 & 05h52m17.12s & -70d55m13.11s & TOI 1948.01 &  & 0.17 & 3.65 &  & 2459559.261 & 2459559.262 & $1.10\times10^{14}$ \\
        
        TIC 317060587 & F-G & 129.80 & 22h30m02.47s & -75d38m47.62s & TOI-1052 b & 0.05 & 0.26 & 9.14 & 0.09 & 2459205.854 & 2459205.857 & $5.43\times10^{13}$* \\
        
        TIC 33911302 & K & 415.55 & 23h43m22.80s & -23d59m52.88s & TOI 341.01 &  & 1.03 & 1.81 &  & 2459020.489 & 2459020.497 & $5.57\times10^{14}$* \\
        
        TIC 351601843 & K-M & 61.46 & 20h39m53.09s & -65d26m58.92s & TOI 1075.01 & 0.03 & 0.16 & 0.61 & 0.01 & 2459458.226 & 2459458.235 & $8.61\times10^{12}$ \\
        
        TIC 370133522\tnote{d} & M & 20.37 & 20h27m42.88s & -56d27m44.23s & TOI 1078.01 & $4.20\times10^{-3}$ & 0.11 & 0.52 & 0.01 & 2458977.225 & 2458977.233 & $1.34\times10^{12}$* \\
           &  &  &  &  &  &  &  &  &  & 2459458.162 & 2459458.168 & $9.46\times10^{11}$ \\
            
        TIC 371443216 & A & 170.70 & 09h50m19.22s & -66d06m50.13s & TOI 1924.01 & 3.10 & 1.53 & 2.82 & 0.05 & 2459600.208 & 2459600.223 & $6.64\times10^{13}$ \\
        
        TIC 380589029 & F & 494.18 & 17h55m33.77s & -61d44m50.64s & TOI 1084.01 & 2.26 & 1.40 & 1.35 & 0.02 & 2458997.203 & 2458997.213 & $7.87\times10^{14}$* \\
        
        TIC 38509907 & K & 57.69 & 04h09m31.10s & -62d51m02.36s & TOI 722.01 &  & 0.21 & 15.30 &  & 2459096.426 & 2459096.439 & $1.07\times10^{13}$* \\
        
        TIC 402026209 & G & 267.21 & 23h34m15.10s & -42d03m42.40s & TOI 232.01 & 1.19 & 1.32 & 1.34 & 0.02 & 2459186.929 & 2459186.934 & $2.30\times10^{14}$* \\
        
        TIC 440887364\tnote{e} & K & 27.50 & 15h00m19.17s & -24d27m15.13s & TOI 836.01 & 0.03 & 0.23 & 8.60 & 0.08 & 2458965.111 & 2458965.121 & $2.44\times10^{12}$* \\
            &   &   &   &   & TOI 836.02 & 0.01 & 0.15 & 3.82 & 0.04 &  &  & \\
      
        TIC 70524163 & G & 327.93 & 02h12m36.97s & -35d23m27.30s & TOI 2421.01 &  & 0.96 & 4.35 &  & 2459559.042 & 2459559.045 & $2.45\times10^{14}$ \\
        
        TIC 80166433\tnote{f} & A & 237.59 & 19h53m09.05s & -46d22m11.97s & TOI 1125.01 &  & 0.37 & 0.77 &  & 2459143.889 & 2459143.900 & $1.82\times10^{14}$* \\
          &  &  &  &  &  &  &  &  &  & 2459147.027 & 2459147.028 & $1.82\times10^{14}$* \\
        \bottomrule
        \end{tabular}
        \begin{tablenotes}[hang]
        \item[a] Each exoplanet was considered to have been observed at a calculated point of secondary ingress or egress (Event Epoch) if captured within a 30 minute window (0.00208 JD) of the observation start time (Obs Epoch). This is in line with the typical 30 minute observing sessions ran by BL during recording.
        \item[b] While TIC 251852984 was observed during the calculated secondary transit window for exoplanet TOI 323.01, the data files associated with this observation were found to be corrupt upon analysis and we were unable to calculate an EIRP metric.
        \item[c] Four TOIs (1233.01 -- 1233.04) associated with TIC 260647166 were observed during the calculated secondary transit window in the same observing session.
        \item[d] TOI 1078.01 was observed to be within a calculated event window in two separate observing sessions.
        \item[e] Exoplanets TOI 836.01 and 836.02 were observed during their respective calculated secondary transit window in the same observing session.
        \item[f] As above, exoplanet TOI 1125.01 was observed to be within a calculated event window in two separate observing sessions.
        \item [*] Due to equipment failure, $EIRP_{min}$ values marked with an asterisk were only recorded over one polarisation and were calculated accordingly.
        \end{tablenotes}
    \end{threeparttable}    
\end{table}
\end{landscape}

\appendix

\end{document}